\newcommand\finalversion

\documentclass{llncs}

\usepackage[latin1]{inputenc}
\usepackage[english]{babel}
\usepackage{verbatim}
\usepackage{tikz} 
\usetikzlibrary{shapes.geometric}
\usetikzlibrary{arrows,positioning,calc,fit}

\usepackage{graphicx}
\DeclareGraphicsExtensions{.jpg, .pdf, .ps}
\usepackage{xcolor}
\usepackage{subfig}

\usepackage{array}

\usepackage{latexsym}
\usepackage{amsmath}
\usepackage{enumitem}

\usepackage{multirow}

\usepackage[normalem]{ulem}
\usepackage{makecell}

\usepackage{wrapfig}

\usepackage{url}
\usepackage[pdftex,bookmarks=true,colorlinks=true,linkcolor=blue,citecolor=blue,urlcolor=blue]{hyperref}

\newcommand{\nat}{\mathtt{nat}}
\newcommand{\real}{\mathtt{real}}

\newcommand{\bool}{\mathtt{bool}}

\newcommand{\delete}[1]{}

\newcommand{\linesTemporal}{788\ }

\newcommand{\linesRefinement}{1209\ }

\newcommand{\linesRefinementReactive}{1144\ }

\newcommand{\linesSimulink}{873\ }

\newcommand{\linesSimplifyRCRS}{2175\ }

\newcommand{\fileTotal}{22\ }
\newcommand{\lineTotal}{27588\ }

\newcommand{\synsop}{\texttt{ o }}
\newcommand{\synpop}{\texttt{ ** }}
\newcommand{\synfb}{\mathtt{feedback}}
\newcommand{\id}{\mathtt{Id}}

\usepackage{listings}
\makeatletter 
\lst@AddToHook{OnEmptyLine}{\vspace{\dimexpr-\baselineskip+\smallskipamount}} 
\makeatother 
\usepackage{textcomp}

\usepackage[firstpage]{draftwatermark}
\SetWatermarkText{\hspace*{4.6in}\raisebox{6.4in}{\includegraphics[scale=0.1]{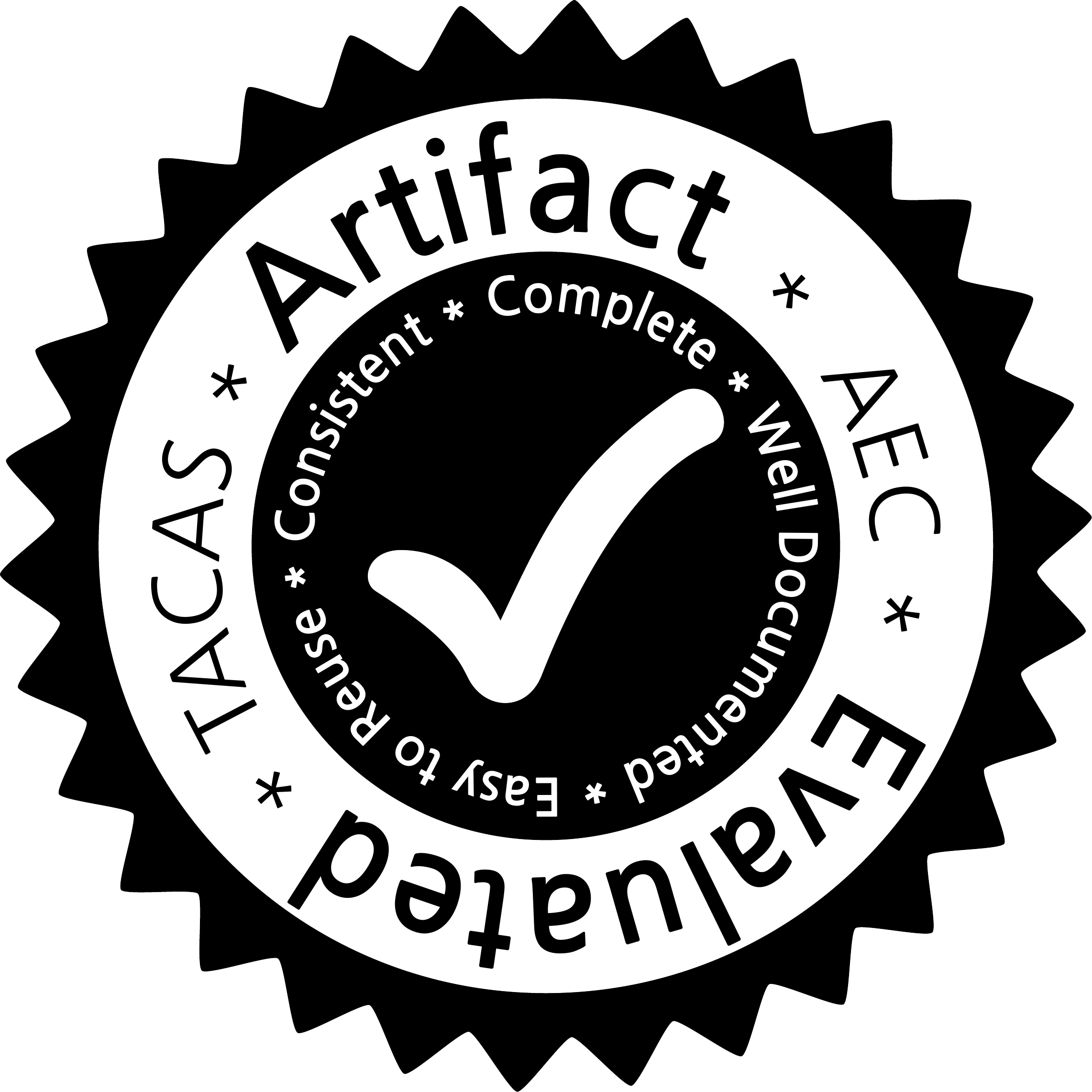}}}
\SetWatermarkAngle{0}

\title{The Refinement Calculus of Reactive Systems Toolset\thanks{This work has been supported by the Academy of Finland and the U.S. National Science Foundation (awards \#1329759 and \#1139138).
\ifdefined\finalversion
The first author was partially supported by the H2020 Programme SRC ESROCOS 
and ERGO 
projects.
\fi
}}

\author{Iulia Dragomir\inst{1} \and Viorel Preoteasa\inst{2} \and Stavros Tripakis\inst{2,3}}

\institute{
\ifdefined\finalversion
Univ. Grenoble Alpes, CNRS, Grenoble INP\thanks{Institute of Engineering Univ. Grenoble Alpes},
\fi
VERIMAG, France \and
Aalto University, Finland \and
University of California, Berkeley, USA}

\begin{document}

\maketitle

\begin{abstract}
We present the Refinement Calculus of Reactive Systems Toolset,
an environment for compositional modeling and reasoning about
reactive systems, built on top of Isabelle, Simulink, and Python.
\end{abstract}

\section{Introduction}
\label{sec:intro}

The {\em Refinement Calculus of Reactive Systems} (RCRS) is a compositional
framework for modeling and reasoning about reactive systems. RCRS
has been inspired by component-based frameworks such as interface automata~\cite{AlfaroHenzingerFSE01} and has its origins in the theory of relational interfaces~\cite{TripakisLHL11}. 
The theory of RCRS has been introduced in~\cite{PreoteasaT14} and is thoroughly described in~\cite{PDT17}.

\begin{figure}[h!]
 \centering
 \includegraphics[width=\textwidth]{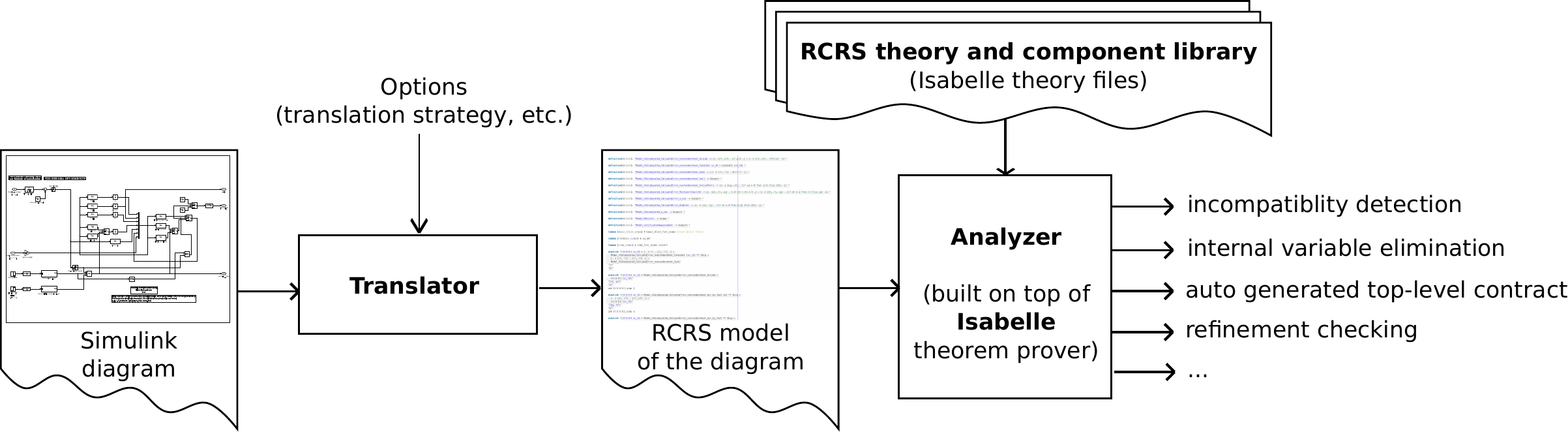}
 \caption{The RCRS toolset.}
 \label{fig_toolset} 
\end{figure}

RCRS comes with a publicly available toolset, the {\em RCRS toolset}
(Fig.~\ref{fig_toolset}), which consists of:
\begin{itemize}
\item A full implementation of RCRS in the Isabelle proof assistant~\cite{NipkowPW02}. 
\item A set of analysis procedures for RCRS components, implemented on top of Isabelle and collectively called the {\em Analyzer}.
\item A {\em Translator} of Simulink diagrams into RCRS code.
\item A {\em library} of basic RCRS components, including a set of
	basic Simulink blocks modeled in RCRS.
\end{itemize} 

An extended version of this paper contains an additional six-page appendix describing a demo of the RCRS toolset~\cite{RCRS_Toolset_arxiv_2017}. 
The extended paper can also be found in a figshare repository~\cite{RCRS_Toolset_figshare}. The figshare repository contains all data (code and models) required to reproduce all results of this paper as well as of~\cite{RCRS_Toolset_arxiv_2017}: see Section~\ref{sec_figshare} for more details.
The RCRS toolset can be downloaded also from the RCRS web page:
\url{http://rcrs.cs.aalto.fi/}.

\section{Modeling Systems in RCRS}
\label{sec:rcrs}

RCRS provides a language of {\em components} to model systems in a modular
fashion. Components can be either {\em atomic} or {\em composite}.
Here are some examples of atomic RCRS components:

\begin{lstlisting}
definition "$\id$ = [: x $\leadsto$ y . y = x :]"
definition "Add =  [: (x, y) $\leadsto$ z . z = x + y :]"
definition "Constant c = [: x::unit $\leadsto$ y . y = c :]"
definition "UnitDelay = [: (x,s) $\leadsto$ (y,s') . y = s $\land$ s' = x :]"
definition "SqrRoot = {. x . x $\ge$ 0 .} $\synsop$ [- x $\leadsto$ $\sqrt{\text{x}}$ -]"
definition "NonDetSqrt = {. x . x $\ge$ 0 .} $\synsop$ [: x $\leadsto$ y . y $\ge$ 0 :]"
definition "ReceptiveSqrt = [: x $\leadsto$ y . x $\ge$ 0 $\longrightarrow$ y = $\sqrt{\text{x}}$ :]"
definition "A = {. x . $\Box\Diamond$x .} $\synsop$ [: x $\leadsto$ y . $\Box\Diamond$y :]"
\end{lstlisting} %

\texttt{Id} models the identity function: it takes input $x$ and returns $y$
such that $y=x$. \texttt{Add} returns the sum of its two inputs.
\texttt{Constant} is parameterized by \texttt{c}, takes no input
(equivalent to saying that its input variable is of type \texttt{unit}),
and returns an output which is always equal to \texttt{c}.
\texttt{UnitDelay} is a {\em stateful} component: \texttt{s} is the
current-state variable and \texttt{s'} is the next-state variable.
\texttt{SqrRoot} is a {\em non-input-receptive} component: its input \texttt{x}
is required to satisfy \texttt{x$\ge$0}. (\texttt{SqrRoot} may be considered
non-atomic as it
is defined as the serial composition of two predicate transformers --
see Section~\ref{sec:isabelle}.)
\texttt{NonDetSqrt} is a {\em non-deterministic} version of \texttt{SqrRoot}:
it returns an arbitrary (but non-negative) \texttt{y}, and not necessarily the
square-root of \texttt{x}.
\texttt{ReceptiveSqrt} is an input-receptive version of \texttt{SqrRoot}:
it accepts negative inputs, but may return an arbitrary output for such inputs.
RCRS also allows to describe components using the temporal logic QLTL, an
extension of LTL with quantifiers~\cite{PDT17}.
An example is component \texttt{A} above. \texttt{A} accepts an infinite
input sequence of \texttt{x}'s, provided \texttt{x} is infinitely often true,
and returns a (non-deterministic) output sequence which satisfies the same
property.

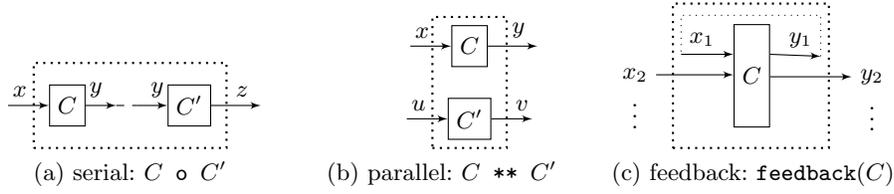
\begin{figure}[t]
  \centering
          \subfloat[serial: $C\synsop C'$]{
                \begin{tikzpicture}
                  \node[draw, minimum height = 4ex] (a) {$C$};
                  \node[left=4ex of a.west] (px) {};
                  \node[shape=coordinate, right = 3ex of a.east] (py) {};
                  \node[draw, minimum height = 4ex, right = 8ex of a] (b) {$C'$};
                  \node[right=4ex of b.east] (pv) {};
                  \node[shape=coordinate, left = 3ex of b.west] (pu) {};
                  \draw[-latex'] (px) --++ (2ex, 0) node[anchor=south] {$x$} -- (px -| a.west);
                  \draw[-latex'] (b.east) --++ (3ex, 0) node[anchor=south] {$z$} -- (b.east -| pv);
                  \draw[-latex'] (a.east) --++ (1ex, 0) node[anchor=south] {$y$} -- (py);
                  \draw[-latex'] (pu) --++ (2ex, 0) node[anchor=south] {$y$} -- (b.west);
                  \draw[dashed] (py) -- (pu);
                  \node[shape=coordinate, above left = 0.1ex and 0.7ex of a](pa){};
                  \node[shape=coordinate, below right = 0.1ex and 0.7ex of b, inner sep = 0](pb){};
                  \node[draw, style=dotted, thick, inner ysep = 2ex, fit = (a)(b)(pa)(pb)](main){};
                \end{tikzpicture}
                \label{fig:serial_comp}
        }
        \qquad
        \subfloat[parallel: $C\synpop C'$]{
                \hspace*{18pt}
                \begin{tikzpicture}
                  \node[draw, minimum height = 4ex] (a) {$C$};
                  \node[draw, minimum height = 4ex, below = 3ex of a] (b) {$C'$};
                  \node[left=4ex of a.west] (x) {};
                  \node[right=4ex of a.east] (y) {};
                  \node[left=4ex of b.west] (u) {};
                  \node[right=4ex of b.east] (v) {};
                  \draw[-latex'] (x) --++ (2ex, 0) node[anchor=south] {$x$} -- (x -| a.west);
                  \draw[-latex'] (u) --++ (2ex, 0) node[anchor=south] {$u$} -- (u -| b.west);
                  \draw[-latex'] (a.east) --++ (3ex, 0) node[anchor=south] {$y$} -- (a.east -| y);
                  \draw[-latex'] (b.east) --++ (3ex, 0) node[anchor=south] {$v$} -- (b.east -| v.west);
                  \node[shape=coordinate, left = 0.7ex of b](pa){};
                  \node[shape=coordinate, right = 0.7ex of b, inner sep = 0](pb){};
                  \node[draw, style=dotted, thick, fit = (a)(b)(pa)(pb)](main){};
                \end{tikzpicture}
                \label{fig:parallel_comp}
        }
        \qquad
        \subfloat[feedback: $\synfb(C)$]{
                \begin{tikzpicture}
                  \node[draw, minimum height = 10ex, minimum width=3ex] (a) {$C$};
                  \node[left = 1ex of a.north west, anchor = north east](x){$x_1$};
                  \draw[-latex'](x.south west) -- (x.south west-|a.west);
                  \node[right = 1ex of a.north east, anchor = north west](y){$y_1$};
                  \draw[latex'-](y.south east) -- (x.south west-|a.east);
                  \draw[dotted](y.south east) -- ++(0,4ex) coordinate(A) -- (A -| x.west) -- (x.south west);
                  \coordinate[left = 4.5ex of x](X);
                  \node[below = 2ex of X](xa){$x_2$};
                  \node[below = -1ex of xa](xb){$\vdots$};
                  \draw[-latex'](xa) -- (xa -| a.west);

                  \coordinate[right = 5ex of y](Y);
                  \node[below = 2ex of Y](ya){$y_2$};
                  \node[below = -1ex of ya](yb){$\vdots$};
                  \draw[-latex'](ya -| a.east) -- (ya);

                  \node[draw, style=dotted,inner ysep = 2ex, thick, fit = (x)(y)(a)]{};

                \end{tikzpicture}
                \label{fig:fb_comp}
        }
  \caption{The three composition operators of RCRS.}
  \label{fig:comp_ops} 
\end{figure}

Composite components are formed by composing other (atomic or composite)
components using three primitive composition operators,
as illustrated in Fig.~\ref{fig:comp_ops}:
$C\synsop C'$ (in series) connects outputs of $C$ to inputs of $C'$;
$C\synpop C'$ (in parallel) ``stacks'' $C$ and $C'$ ``on top of each other'';
and $\synfb(C)$ connects the first output of $C$ to its first input.
These operators are sufficient to express any block diagram, as described
in Section~\ref{sec:translator}.

\section{The Implementation of RCRS in Isabelle}
\label{sec:isabelle}

RCRS is fully implemented in the Isabelle theorem prover.
The RCRS implementation currently consists of \fileTotal Isabelle {\em theories}
(\texttt{.thy} files), totalling \lineTotal lines of Isabelle code. 
Some of the main theories are described next.

Theory \texttt{Refinement.thy} (\linesRefinement lines) contains a standard implementation of refinement calculus \cite{backwright:98}. Systems are modeled as monotonic predicate transformers
\cite{dijkstra:75} with a weakest precondition interpretation.
Within this theory we implemented non-deterministic and deterministic update statements, assert statements, parallel composition, refinement  and other operations, and proved necessary properties of these.

Theory \texttt{RefinementReactive.thy} (\linesRefinementReactive lines) extends \texttt{Reactive.thy} to reactive systems by introducing predicates over infinite traces in addition to predicates over values, 
and {\em property} transformers in addition to predicate transformers~\cite{PreoteasaT14,PDT17}. 

Theory \texttt{Temporal.thy} (\linesTemporal lines) implements a semantic version of QLTL, where temporal operators are interpreted as predicate transformers.
For example, the operator $\Box$, when applied to the predicate on infinite
traces $(x>0): (\nat \to \real) \to \bool$, returns another predicate
on infinite traces $\Box(x>0): (\nat \to \real) \to \bool$.
Temporal operators have been implemented to be polymorphic in the sense that
they apply to predicates over an arbitrary number of variables.

Theory \texttt{Simulink.thy} (\linesSimulink lines) defines a subset of the 
basic blocks in the Simulink library as RCRS components
(at the time of writing, 48 Simulink block types can be handled).
In addition to discrete-time, we can handle continuous-time blocks with
a fixed-step forward Euler integration scheme. For example, Simulink's 
integrator block can be defined in two equivalent ways as follows:
\begin{lstlisting}
definition "Integrator dt = [- (x,s) $\leadsto$ (s, s+x*dt) -]"
definition "Integrator dt = [: (x,s) $\leadsto$ (y,s'). y=s $\land$ s'=s+x*dt :]"
\end{lstlisting} %
The syntax \texttt{[- x $\leadsto f$(x) -]} assumes that $f$ is a function,
whereas \texttt{[: :]} can be used also for relations (i.e., non-deterministic systems). Using the former instead of the latter to describe deterministic
systems aids the Analyzer to perform simplifications -- see Section~\ref{sec:analyzer}.

Theory \texttt{SimplifyRCRS.thy} (\linesSimplifyRCRS lines) implements
several of the Analyzer's procedures. In particular, it contains
a simplification procedure which reduces composite RCRS components into atomic
ones (see Section~\ref{sec:analyzer}).

In addition to the above, there are several theories containing a proof
of correctness of our block-diagram translation strategies 
(see Section~\ref{sec:translator} and~\cite{PDT:2016}), 
dealing with Simulink types~\cite{DragomirPreoteasaTripakisFORTE2017},
generating Python simulation code, and many more.
A detailed description of all these theories and graphs depicting their dependencies is included in the documentation of the toolset.

The syntax of RCRS components is implemented in Isabelle using a {\em shallow embedding} \cite{Boulton:1992}. This has the advantage of all datatypes and 
other mechanisms of Isabelle (e.g., renaming) being available for 
component specification, but also the disadvantage of not being
able to express properties and simplifications of the RCRS language within Isabelle, as discussed in~\cite{PDT17}.
A {\em deep embedding}, in which the syntax of components is defined as a datatype of Isabelle, is possible, and is left as an open future work direction.

\section{The Translator}
\label{sec:translator}

The Translator, called \texttt{simulink2isabelle}, translates {\em hierarchical block diagrams} (HBDs), and in particular Simulink models, into RCRS
theories~\cite{DragomirPT16}.
The Translator (implemented in about 7100 lines of Python code)
takes as input a Simulink model (\texttt{.slx} file) and a list of options and generates as output an Isabelle theory (\texttt{.thy} file). The output file contains:
(1) the definition of all instances of basic blocks in the Simulink diagram (e.g., all Adders, Integrators, Constants, etc.) as atomic RCRS components; 
(2) the bottom-up definition of all subdiagrams as composite RCRS components; 
(3) calls to simplification procedures; 
and
(4) theorems stating that the resulting simplified components are equivalent to the original ones.
The \texttt{.thy} file may also contain additional content depending on
user options as explained below.

As shown in~\cite{DragomirPT16}, there are many possible ways to translate
a block diagram into an algebra of components with the three primitive
composition operators of RCRS. This means that step (2) above is not unique.
\texttt{simulink2isabelle} implements the several translation strategies
proposed in~\cite{DragomirPT16} as user options.
\begin{wrapfigure}{R}{0.4\textwidth} 
  \includegraphics[scale=0.55]{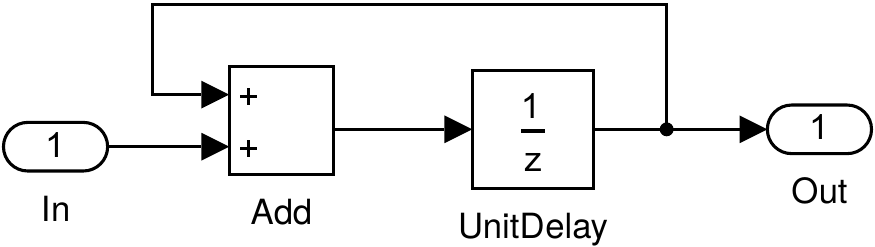}
  \caption{A Simulink diagram.}
  \label{fig:diagram} 
\end{wrapfigure}
For example, when run on the Simulink diagram of Fig.~\ref{fig:diagram},
the Translator produces a file similar to the one shown in Fig.~\ref{fig_translation}.
\texttt{IC\_Model} and \texttt{FP\_Model} are composite RCRS components generated automatically w.r.t. two different translation strategies, implemented by user options \texttt{-ic} and \texttt{-fp}.
The \texttt{simplify\_RCRS} construct is explained in Section~\ref{sec:analyzer} that
follows.

Other user options to the Translator include: 
whether to flatten the input diagram,
optional typing information for wires,
and whether to generate in addition to the top-level STS component, a QLTL
component representing the temporal behavior of the system.
The user can also ask the Translator to generate:
(1) components w.r.t. all
translation strategies;
(2) the corresponding theorems showing that these
components are all semantically equivalent;
and 
(3) Python simulation scripts for the top-level component.

\begin{figure}[t]
\begin{lstlisting}

theory Summation imports ... 
begin 

named_theorems basic_simps
lemmas basic_simps = simulink_simps 

definition [basic_simps]: "Split  = [- a $\leadsto$ a, a -]"
definition [basic_simps]: "Add  = [- f, g $\leadsto$ f + g -]"
definition [basic_simps]: "UnitDelay  = [- d, s $\leadsto$ s, d -]" 

simplify_RCRS "IC_Model = $\synfb$([- f, g, s $\leadsto$ (f, g), s -] $\synsop$ 
    (Add $\synpop$ Id) $\synsop$ UnitDelay $\synsop$ (Split $\synpop$ Id) $\synsop$ 
    [- (f, h), s$'$ $\leadsto$ f, h, s$'$ -])"
    "(g, s)" "(h, s$'$)"
  
simplify_RCRS "FP_Model = $\synfb$ ($\synfb$ ($\synfb$ ([- f, d, a, g, s 
   $\leadsto$ (f, g), (d, s), a -] $\synsop$ (Add $\synpop$ UnitDelay $\synpop$ Split) $\synsop$ 
   [- d, (a, s$'$), (f, h) $\leadsto$  f, d, a, h, s$'$ -])))"
   "(g, s)" "(h, s$'$)"
end 
\end{lstlisting} 
\caption{Auto-generated Isabelle theory for the Simulink diagram of Fig.~\ref{fig:diagram}.}\label{fig_translation} 
\end{figure}

\section{The Analyzer}
\label{sec:analyzer}

\begin{sloppypar}
The Analyzer is a set of procedures implemented on top of Isabelle and
ML, the programming language of Isabelle.
These procedures implement a set of functionalities such as simplification,
compatibility checking, refinement checking, etc.
Here we describe the main functionalities, implemented by the 
\texttt{simplify\_RCRS}
construct. 
As illustrated in Fig.~\ref{fig_translation},
the general usage of this construct is
\texttt{simplify\_RCRS "Model = C" "in" "out"}, where \texttt{C} is a (generally composite)
component and \texttt{in}, \texttt{out} are (tuples of) names for its input and output variables.
When such a statement is executed in Isabelle, it performs the following
steps:
(1) It creates the definition \texttt{Model = C}.
(2) It {\em expands} \texttt{C}, meaning that it replaces all atomic components
and all composition operators in \texttt{C} with their definitions. This results
in an Isabelle expression \texttt{E}.
\texttt{E} is generally a complicated expression, containing formulas
with quantifiers, \texttt{case} expressions for tuples, function compositions, 
and several other operators.
(3) \texttt{simplify\_RCRS} {\em simplifies} \texttt{E}, by eliminating quantifiers, renaming variables, and performing several other simplifications.
The simplified expression, \texttt{F}, is of the form \texttt{\{.$p$.\} o [:$r$:]}, where $p$ is a predicate on input variables and $r$ is a relation on input and output variables.
That is, \texttt{F} is an atomic RCRS component.
(4) \texttt{simplify\_RCRS} generates a theorem stating that \texttt{Model} is semantically equivalent to \texttt{F},
and also the mechanized proof of this theorem (in Isabelle).
Note that the execution by the Analyzer of the \texttt{.thy} file generated by the Translator is fully automatic, despite the fact that Isabelle generally requires human interaction. 
This is thanks to the fact that the theory generated by the Translator contains all declarations (equalities, rewriting rules, etc.) neccessary for the Analyzer to produce the simplifications and their mechanical proofs, without user interaction.
\end{sloppypar}

For example, when the theory in Fig.~\ref{fig_translation} is executed,
the following theorem 
is generated and proved automatically:
\begin{lstlisting}[xleftmargin=.3\textwidth]
Model = [- (g, s) $\leadsto$ (s, s+g) -]
\end{lstlisting} %
where \texttt{Model} is either \texttt{IC\_Model} or \texttt{FP\_Model}.
The rightmost expression is the automatically generated 
simplification of the top-level system to an atomic RCRS component.

If the model contains {\em incompatibilities}, where for instance the input
condition of a block like \texttt{SqrRoot} cannot be guaranteed by the upstream
diagram, the top-level component automatically simplifies to $\bot$ (i.e.,
false). Thus, in this usage scenario, RCRS can be seen as a static analysis and behavioral type checking and inference tool for Simulink.

\section{Case Study}
\label{sec:eval}

We have used the RCRS toolset on several case studies, the most significant
of which is a real-world benchmark provided by Toyota~\cite{JinDKUB14b}.
The benchmark consists of a set of Simulink diagrams
modeling a Fuel Control System.\footnote{
We downloaded the Simulink models from
\url{https://cps-vo.org/group/ARCH/benchmarks}.
One of those models is made available in the figshare 
repository~\cite{RCRS_Toolset_figshare} --
see also Section~\ref{sec_figshare}.
}
A typical diagram in the above suite contains 3 levels of hierarchy, 104 
Simulink blocks in total (out of which 8 subsystems), and 101 wires 
(out of which 8 are feedbacks, the most complex composition operator in RCRS).
Using the Translator on this diagram results in a \texttt{.thy} file of 
1671 lines and 57037 characters. Translation time is negligible.
The Analyzer simplifies this model to a top-level atomic STS component
with no inputs, 7 (external) outputs and 14 state variables (note that
all internal wires have been automatically eliminated in this top-level description). 
Simplification takes approximately 15 seconds and generates a
formula which is 8337 characters long.
The formula is consistent (not false), which proves
statically that the original Simulink diagram has no incompatibilities.
More details about the case study can be found in~\cite{DragomirPT16,RCRS_Toolset_arxiv_2017}.

\section{Data Availability Statement}
\label{sec_figshare}

All results mentioned in this paper as well as in the extended version of
this paper~\cite{RCRS_Toolset_arxiv_2017} are
fully reproducible using the code, data, and instructions available
in the figshare repository: \url{https://doi.org/10.6084/m9.figshare.5900911}.

The figshare repository contains the full implementation of the RCRS toolset,
including the formalization of RCRS in Isabelle, the Analyzer, the RCRS
Simulink library, and the Translator.
The figshare repository also contains sample Simulink models, including
the Toyota model discussed in Section~\ref{sec:eval}, a demo file
named \texttt{RCRS\_Demo.thy}, and detailed step-by-step instructions on 
how to conduct a demonstration and how to reproduce the results of this paper. Documentation on RCRS is also provided.

The figshare repository provides a snapshot of RCRS as of February 2018.
Further developments of RCRS will be reflected on the RCRS web page:
\url{http://rcrs.cs.aalto.fi/}.

\appendix
\section{Demo}
\label{sec:demo}

\subsection{Basic Reasoning in RCRS}

\begin{figure}[t]
  \centering
  \subfloat[A square root system.]{\includegraphics[scale=0.25]{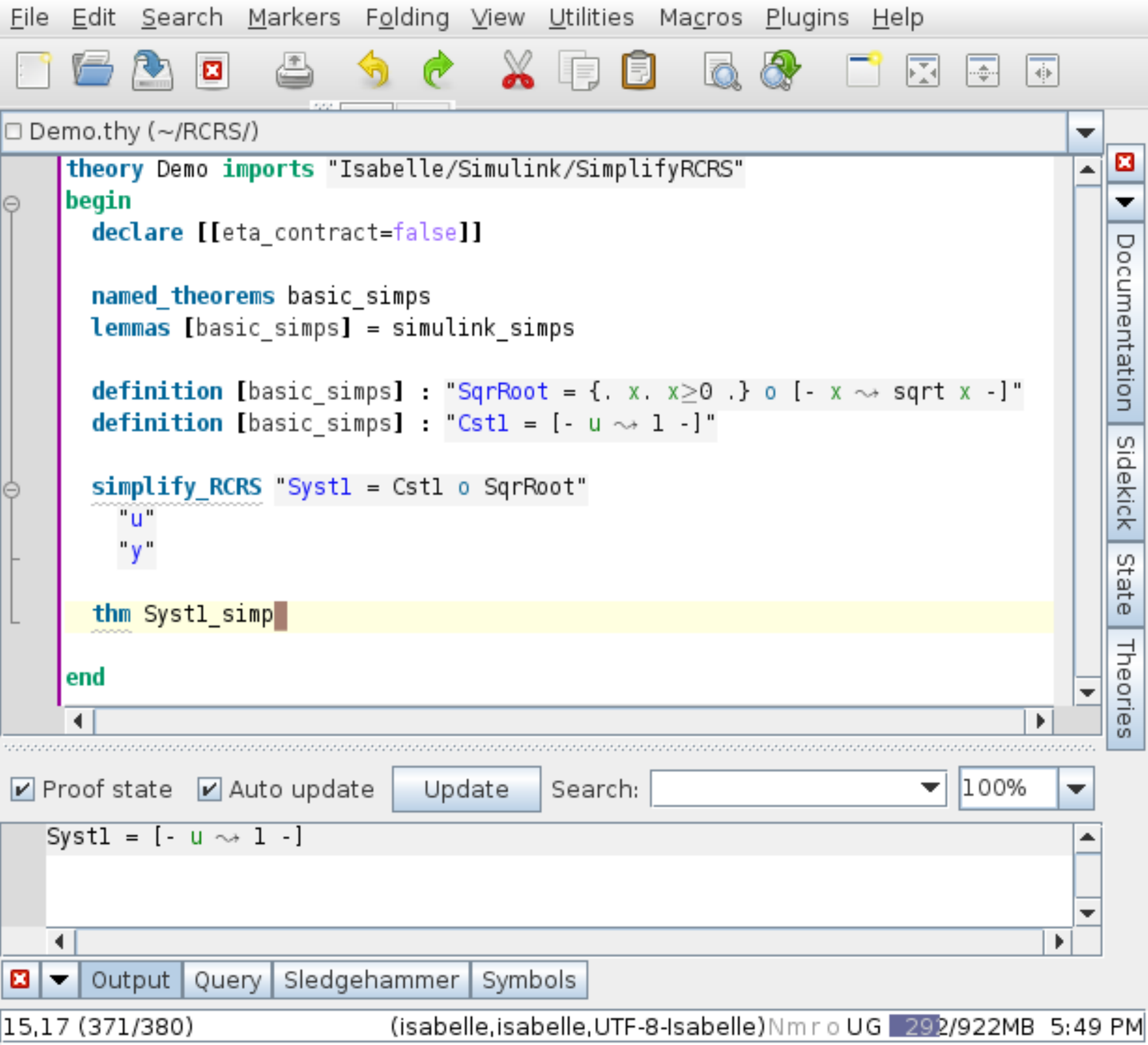} \label{fig:comp_syst}} \qquad
  \subfloat[Incompatibility detected.]{\includegraphics[scale=0.25]{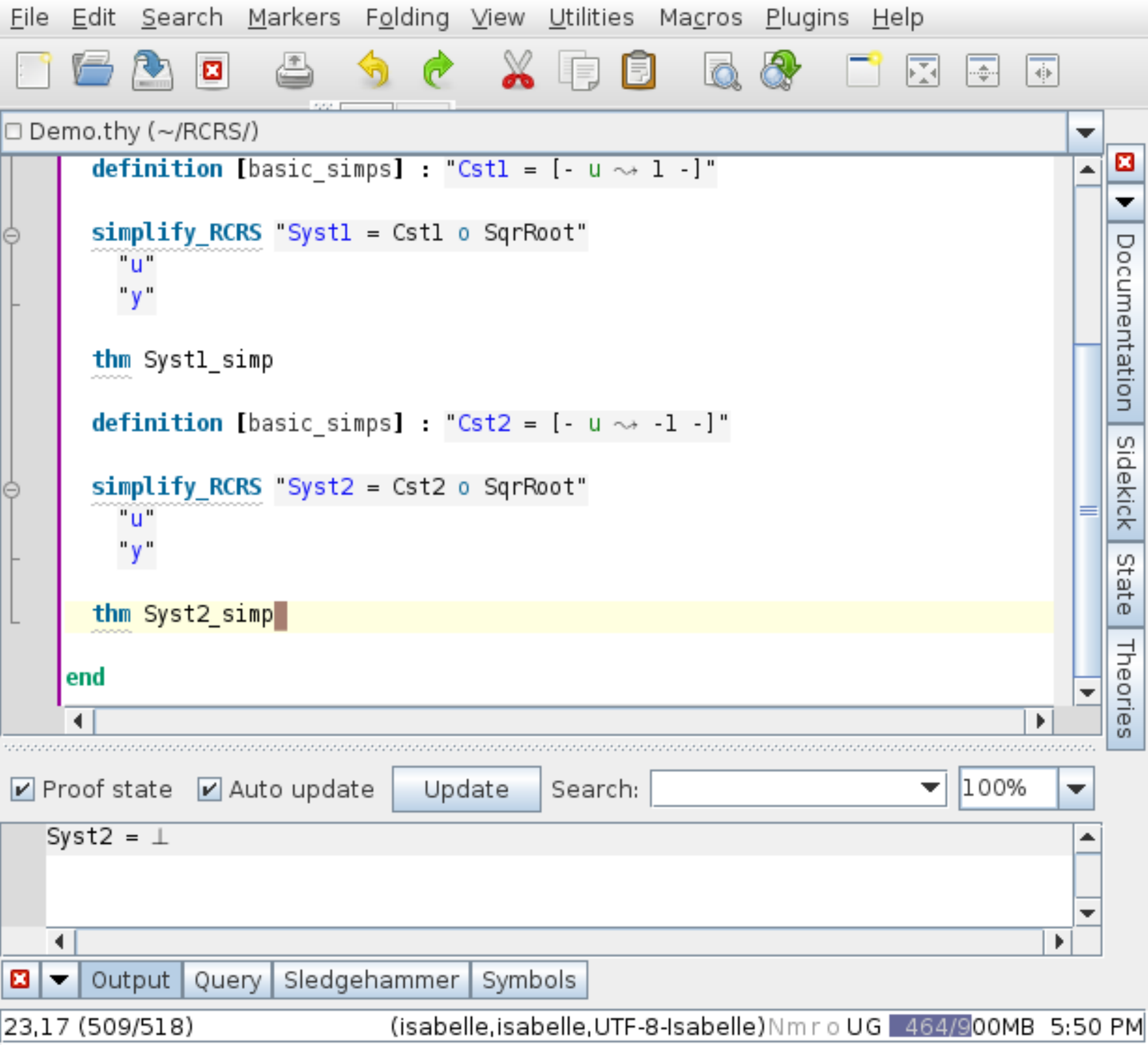} \label{fig:incomp_syst}}
  \caption{A system in RCRS Isabelle and detection of incompatibility.} \vspace{-4ex}
\end{figure}

We begin by showing how to perform some basic reasoning in RCRS.
We open Isabelle and create a new theory file \texttt{RCRS\_Demo.thy} with
initial skeleton as shown below (to import the RCRS Isabelle theories,
and to declare the collection of theorems and lemmas that we will use later
for simplification):
\begin{lstlisting}
theory RCRS_Demo imports "Isabelle/Simulink/SimplifyRCRS"
begin
  named_theorems basic_simps
  lemmas [basic_simps] = simulink_simps
end
\end{lstlisting}

We next define three RCRS components (Fig.~\ref{fig:comp_syst}):
\texttt{SqrRoot} (modeling the square root function, see Section~\ref{sec:rcrs}),
\texttt{Const1} (modeling the constant $1$),
and the composite component \texttt{Syst1}, formed by composing 
\texttt{Const1} and \texttt{SqrRoot} in series.
We explain the notation and point out that \texttt{SqrRoot} is non-input-receptive,
meaning that it rejects negative inputs.

The \texttt{simplify\_RCRS} construct does several things.
First, it defines the composite component \texttt{Syst1}.
Second, it gives names to the external inputs and outputs of \texttt{Syst1}:
\texttt{"u"} and \texttt{"y"} in this case.
Third, it calls the RCRS {\em Analyzer}.
The Analyzer is a set of procedures that we implemented on top of Isabelle,
to perform a number of static analysis tasks. Among these tasks are the
expansion and simplification of the logical formulas involved
in RCRS expressions like the ones here. In this example, the Analyzer
finds that \texttt{Syst1} simplifies to \texttt{[- u $\leadsto$ 1 -]},
as shown at the bottom of the Isabelle window, in the frame called ``Output''.
This result is to be expected, as the whole system outputs the constant $1$.

Let us now see what happens if we replace the constant $1$ by $-1$.
As we can see (Fig.~\ref{fig:incomp_syst}), the Analyzer now returns $\bot$.
In RCRS $\bot$ models the {\em invalid} component, and the fact that a system
simplifies to $\bot$ indicates some kind of inconsistency. The inconsistency
here is obviously that $-1$ violates the input condition of \texttt{SqrRoot}.
So the components \texttt{Const2} and \texttt{SqrRoot} are {\em incompatible}.

So far, all our components were {\em deterministic} systems, in the sense
they map each input to a unique output. Let us continue our example by
showing what happens if we try to connect \texttt{SqrRoot} to a non-deterministic
component which can output any value. 
\begin{lstlisting}
definition [basic_simps] : "true = [: u $\leadsto$ y::real. True]"
\end{lstlisting} 
This component, called \texttt{true}, can be seen as modeling a ``black-box'' system for
which we have no information (e.g., no available source code) or which we
are unable to analyze. Obviously, in such a case, it is difficult to
guarantee anything.
Therefore, connecting \texttt{true} to \texttt{SqrRoot} should result in an 
incompatibility.
Indeed, when we call the Analyzer's simplification procedure:
\begin{lstlisting}
simplify_RCRS  "Syst3 = true $\synsop$ SqrRoot" "u" "y"
thm Syst3_simp
\end{lstlisting}
we see in Isabelle's output window that the system simplifies to
\begin{lstlisting}
Syst3 = {.u. $\forall$y. 0 $\le$ y.} $\synsop$ [: u $\leadsto$ y. $\exists$z. y = sqrt z:]
\end{lstlisting}
The formula $\forall y. 0 \le y$ is unsatisfiable, which means that \texttt{Syst3} is inconsistent, indicating the incompatibility.
Unfortunately, the simplification above does not result in an expression which
is as simple as it can be, which is due to Isabelle's limitations in
simplifying expressions with quantifiers. In this case we have to ``help''
Isabelle, by recognizing that the formula $\forall y. 0 \le y$ is unsatisfiable.
We state this as a lemma:
\begin{lstlisting}
lemma aux1: "($\forall$y::real. 0 $\le$ y) = False"
  (* sledgehammer *)
  using le_minus_one_simps(3) by blast
\end{lstlisting}
and prove it using Isabelle's ``sledgehammer'' mechanism 
(strangely, Isabelle's ``auto'' does not work on this formula, even though
it appears to work on the more complex formula discussed below).
Having proved this lemma, we can call the simplification procedure again
and ask it this time to use the lemma as a fact:
\begin{lstlisting}
simplify_RCRS  "Syst4 = true $\synsop$ SqrRoot" "u" "y"    
  use (aux1)
\end{lstlisting}
This time simplification succeeds and produces:
\begin{lstlisting}
Syst4 = $\bot$
\end{lstlisting}

Now, suppose that we have a component for which
we know something, for instance, that its output $y$ is greater than its
input $x$ plus $1$:
\begin{lstlisting}
definition [basic_simps]: "A = [: x $\leadsto$ y. y$\ge$x+1 :]"
\end{lstlisting}
Let us see what happens if we connect \texttt{A} to \texttt{SqrRoot},
and try to simplify:
\begin{lstlisting}
simplify_RCRS  "Syst5 = A $\synsop$ SqrRoot" "x" "y"    
\end{lstlisting}
We get:
\begin{lstlisting}
Syst5 = {.x.$\forall$y$\ge$x+1. 0$\le$y.} $\synsop$ [:x$\leadsto$y.$\exists$z$\ge$x+1. y=sqrt z:]
\end{lstlisting}
Again, Isabelle has trouble eliminating the quantifiers from the formulas
and needs our help. We recognize that the formula in the precondition is
equivalent to $x\ge -1$, and state this as a lemma:
\begin{lstlisting}
lemma aux2: "($\forall$y::real$\ge$x + 1. 0$\le$y) = (x$\ge$-1)"
  by auto
\end{lstlisting}
(Interestingly, Isabelle manages to prove this result automatically, even though
the formula involved seems more complex than the unsatisfiable formula above.)
We can now use the above lemma to simplify further:
\begin{lstlisting}
simplify_RCRS  "Syst6 = A $\synsop$ SqrRoot" "x" "y"   
  use (aux2)
\end{lstlisting}
and we get:
\begin{lstlisting}
Syst6= {.x. -1 $\le$ x.} $\synsop$ [:x$\leadsto$y.$\exists$z$\ge$x+1. y=sqrt z:]
\end{lstlisting}
The precondition $x\ge -1$ is as simple as it can be, but the
postcondition can be simplified further by eliminating the existential 
quantifier. This requires manual intervention to Isabelle which will be
explained in the demo.
The end result is the following lemma
\begin{lstlisting}
lemma "Syst6 = {. x. -1 $\le$ x .} $\synsop$ [: x $\leadsto$ z. z $\ge$ sqrt (x+1)]"
  apply ...
\end{lstlisting}
which depends on proving 
\begin{lstlisting}
lemma aux3: "($\lambda$ x z. -1$\le$x $\land$ ($\exists$y. y$\ge$(x+1) $\land$ z = sqrt y)) = 
               ($\lambda$ x z. -1$\le$x $\land$ z $\geq$ sqrt (x+1))"
  apply ...
\end{lstlisting}
These proofs will be included in the software artifacts and details will be provided during the demonstration.

The above examples illustrated several of the features of RCRS as a reasoning
tool, similar to a behavioral type checking and inference engine.
Indeed, detecting incompatible connections is akin to catching type errors
in programs, and inferring conditions such as the condition on the
input in the last example above is akin to type inference.
In addition to these capabilities, RCRS can be used to check {\em refinement}
(and its counterpart, {\em abstraction}) between components.
We next show how to prove that \texttt{SqrRoot} refines 
\texttt{NonDetSqrt} and that \texttt{ReceptiveSqrt} refines \texttt{SqrRoot}
(\texttt{NonDetSqrt} and \texttt{ReceptiveSqrt} are defined in Section~\ref{sec:rcrs}).
We have:
\begin{lstlisting}
lemmas [basic_simps] =  comp_rel_simps basic_block_rel_simps update_def refinement_simps  

definition [basic_simps] : "NonDetSqrt = {.x.x$\geq$0.}$\synsop$[: x $\leadsto$ y. y$\geq$0 :]"
  
lemma "NonDetSqrt $\le$ SqrRoot"
  by (auto simp add: basic_simps)
  
definition [basic_simps] : "ReceptiveSqrt = [: x$\leadsto$y. x$\geq$0 $\rightarrow$ y=sqrt x:]"

lemma "SqrRoot $\le$ ReceptiveSqrt" 
  by (simp add: basic_simps) 
\end{lstlisting}
The first line instructs the Analyzer to use simplification rules for relational predicate transformers~\cite{PDT17}.
Then, we define the new components and state the refinements as lemmas.
The proofs are simple and are based on the necessary and sufficient conditions
for checking refinement included in the RCRS library:
\begin{lstlisting}
lemma assert_demonic_refinement: "({.p.} $\synsop$ [:r:] $\le$ {.p'.} $\synsop$ [:r':]) = $\;\;\;$ (p $\le$ p' $\land$ ($\forall$ x . p x $\rightarrow$ r' x $\le$ r x))"
  by  (auto simp add: le_fun_def assert_def demonic_def)
    
lemma spec_demonic_refinement: "({.p.} $\synsop$ [:r:] $\le$ [:r':]) = $\;\;\;\;\;\;\;\;\;\;\;\;\;\;\;\;\;\;\;$ ($\forall$x. p x $\rightarrow$ r' x $\le$ r x)"
  by  (auto simp add: le_fun_def assert_def demonic_def)
\end{lstlisting}

\subsection{The Simulink Translator}

\begin{figure}[!t] 
  \centering
  \begin{minipage}{.45\linewidth}
    \centering
    \subfloat[A Simulink diagram]{\hspace{8ex}\includegraphics[scale=0.6]{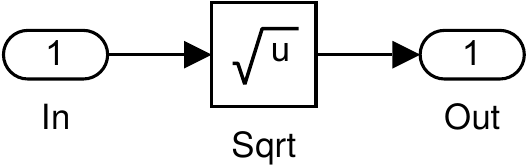}\hspace{8ex}\label{fig:simulink_sqrt}}  \vspace{2ex}
    \subfloat[Running the translator on the Simulink diagram above]{\includegraphics[scale=0.25]{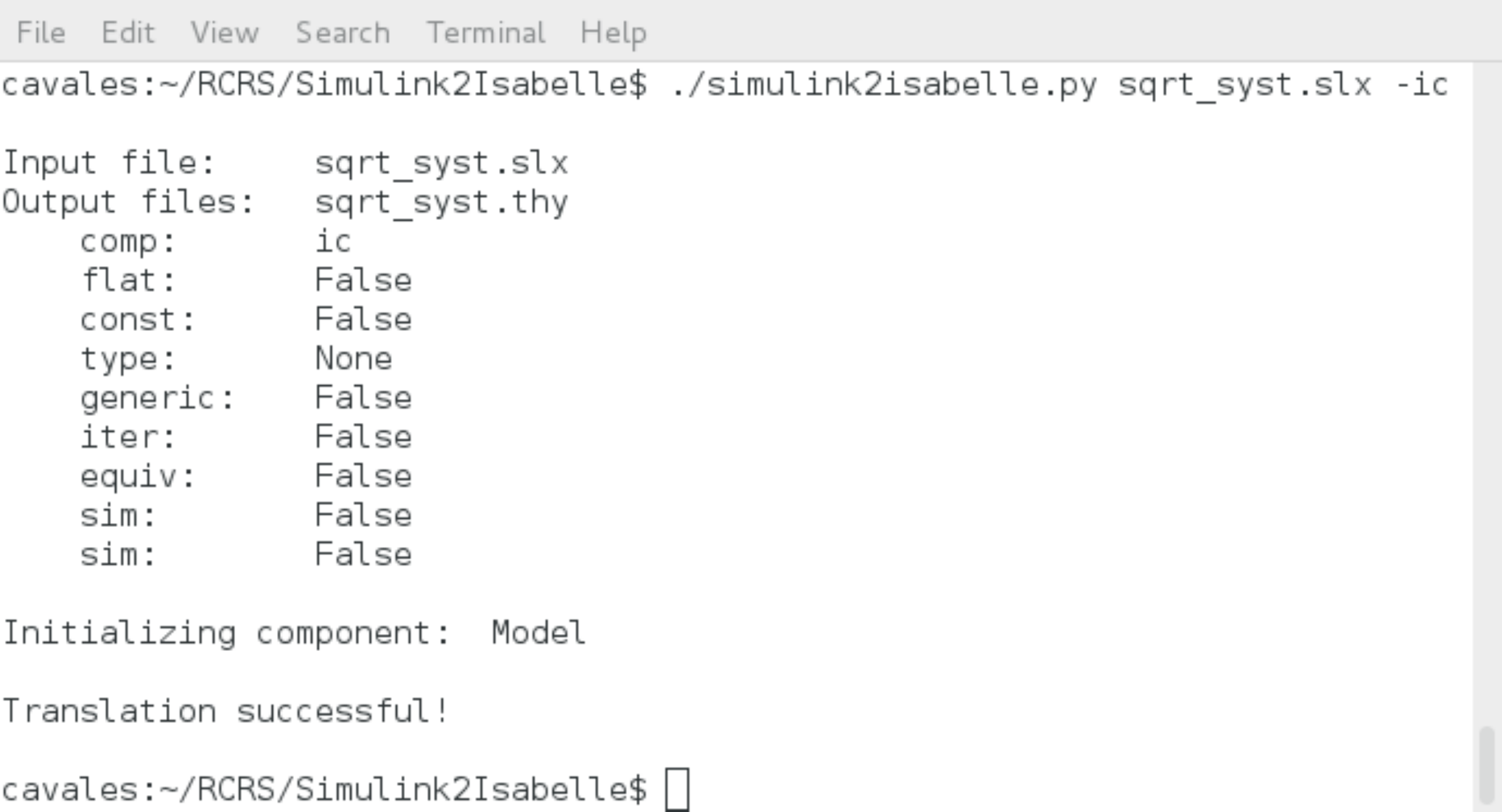}\label{fig:translator_sqrt}}
  \end{minipage} \qquad\quad
  \begin{minipage}{.45\linewidth}
    \subfloat[The generated RCRS theory]{\includegraphics[scale=0.25]{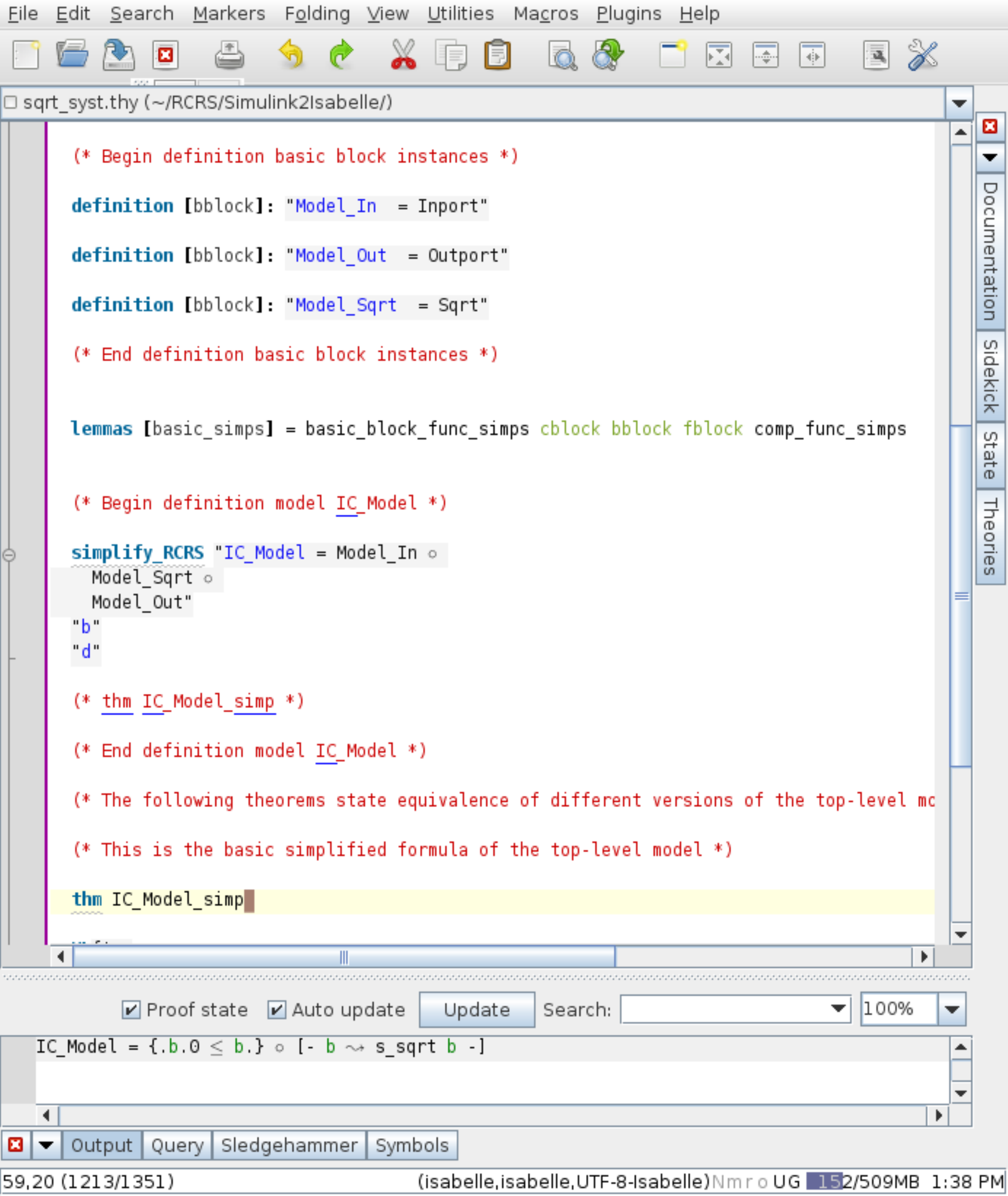}\label{fig:generated_sqrt}}
  \end{minipage}
  \caption{Applying the RCRS toolchain on a Simulink diagram.}
  \label{fig:sqrt} \vspace{-5ex}
\end{figure}

The RCRS toolset also contains a {\em Translator} which takes as input
Simulink diagrams (\texttt{.slx} files) and generates RCRS Isabelle
theories (\texttt{.thy} files). 
The Translator is a Python program called \texttt{simulink2isabelle}.
We illustrate its use by building
a simple Simulink model (Fig.~\ref{fig:simulink_sqrt}) and using 
\texttt{simulink2isabelle} to translate it to RCRS:
\begin{lstlisting}[xleftmargin=.2\textwidth]
./simulink2isabelle.py sqrt_syst.slx -ic
\end{lstlisting}
The execution of the translator is shown in Fig.~\ref{fig:translator_sqrt}, and the generated RCRS Isabelle theory is rendered in Fig.~\ref{fig:generated_sqrt}. 
The option \texttt{-ic} tells the Translator to use a specific algorithm
for translating block diagrams to composite RCRS components. Here we pause
and introduce another feature of RCRS which we haven't talked about so far,
namely, its composition operators. RCRS offers three primitive composition
operators: serial (which we have already seen), parallel, and feedback.
We will illustrate the latter two with examples coming from Simulink.

Before doing so, let us go over the file \texttt{sqrt\_syst.thy} generated by
the Translator. The structure of the file is similar to the files we have
manually created earlier.
One difference is that the definition of the basic components, corresponding
to basic Simulink blocks, refers to predefined RCRS components included
in the \texttt{Simulink.thy} library. We can CTRL-click to see the definition
of these components: we CTRL-click on top of a component which opens
automatically the \texttt{Simulink.thy} file.

We explain briefly the \texttt{Simulink.thy} file, which contains the RCRS
formalization of several Simulink basic blocks.
For example, the \texttt{Gain} block 
\begin{lstlisting}[xleftmargin=.2\textwidth]
definition "Gain k = [- x $\leadsto$ x * k -]"
\end{lstlisting}
is parameterized by $k$, the multiplication factor.
The \texttt{Integrator} block 
\begin{lstlisting}[xleftmargin=.15\textwidth]
definition "Integrator dt = [- x, s $\leadsto$ s, s+x*dt -]"
\end{lstlisting}
is parameterized by $dt$, the time step of the simple Euler integration method
that we use.
Here we take also the opportunity to explain how we model in RCRS {\em stateful}
components: \texttt{s} is the state variable in the \texttt{Integrator}
component above.

\begin{figure}
  \centering
  \includegraphics[scale=0.5]{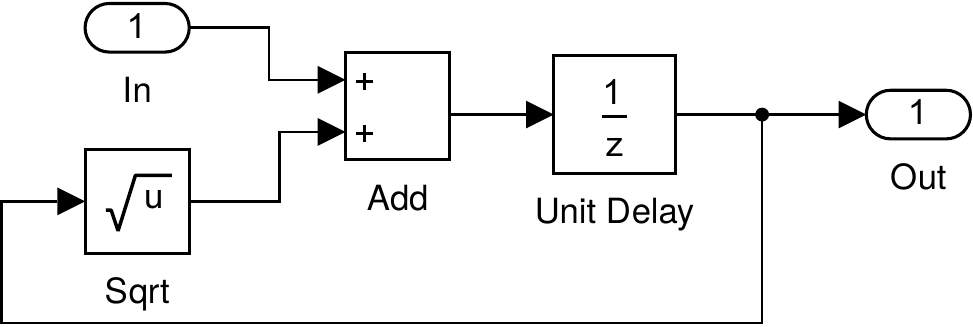}
  \caption{Simulink model \texttt{simple\_syst.slx}.}
  \label{fig:simulink_ss} 
\end{figure}

In all our examples so far we have seen only serial composition, $\synsop$. RCRS contains three primitive composition operators: serial, parallel, and feedback, 
which we illustrate in the sequel.
Consider the Simulink diagram of Fig.~\ref{fig:simulink_ss}.
We execute the Translator
\begin{lstlisting}[xleftmargin=.1\textwidth]
./simulink2isabelle.py simple_syst.slx -ic
\end{lstlisting}
and obtain a theory where 
\texttt{Model\_SqrRoot}, \texttt{Model\_Add} and \texttt{Model\_UnitDelay} are defined respectively as \texttt{SqrRoot}, \texttt{Add} and \texttt{UnitDelay}
above, and the top-level system is defined as shown below:
\delete{
simplify_RCRS "IC_Model = $\synfb$ ([- a, f, si_l $\leadsto$ (f, a), si_l -] $\synsop$
  (Model_In $\synpop$ Model_SqrRoot $\synsop$ Model_Add) $\synpop$ Id $\synsop$
   Model_UnitDelay $\synsop$ Model_Split6 $\synpop$ Id $\synsop$
   [- (a, b), so_l $\leadsto$ a, b, so_l -]) $\synsop$
   Model_Out $\synpop$ Id"
  "(f, si_l)"
  "(j, so_l)" 
}
\begin{lstlisting}
simplify_RCRS "IC_Model = feedback ([- b, f, si_j $\leadsto$ (f, b), si_j -] $\circ$
    (Model_In ** Model_Sqrt $\circ$ Model_Add) ** Skip $\circ$
    Model_UnitDelay $\circ$ Model_Split6 ** Skip $\circ$
    [- (a, b), so_j $\leadsto$ b, a, so_j -]) $\circ$
    Model_Out ** Skip"
  "(f, si_j)"
  "(h, so_j)"
\end{lstlisting}
The parallel and feedback composition operators are denoted $\synpop$ and $\synfb$.
$\synpop$ binds stronger than $\synsop$, so that \texttt{A$\synpop$B$\synsop$C}
is equivalent to \texttt{(A$\synpop$B)$\synsop$C}.
\texttt{Skip} is identical to \texttt{Id}.
The current and next state variables are denoted 
\texttt{si\_j} and \texttt{so\_j}.

\subsection{An Industrial Benchmark}

\begin{figure}
  \centering
  \vspace{-.4cm}
  \includegraphics[scale=0.35]{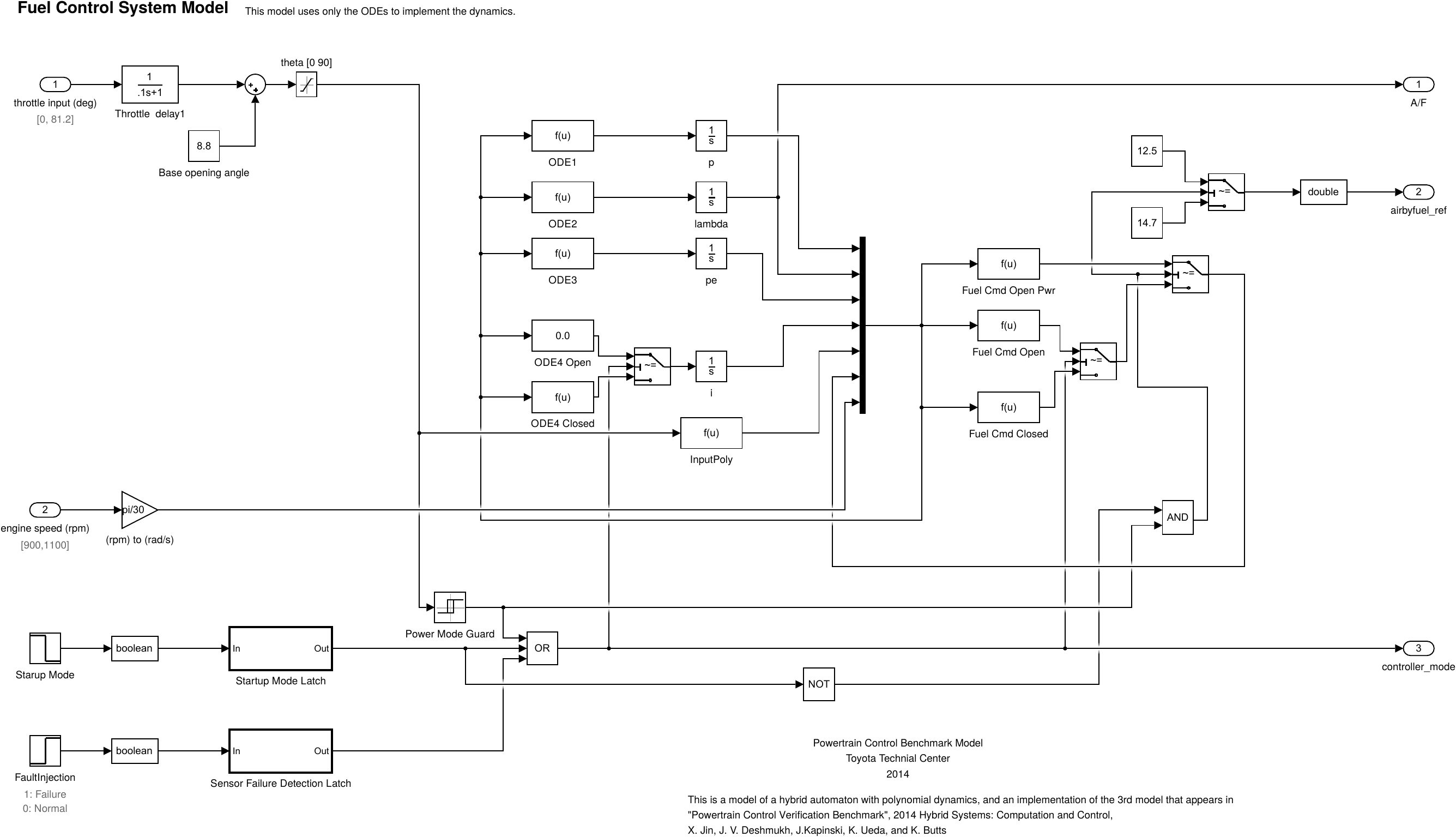}
  \caption{Excerpt of the Toyota Fuel Control System Simulink model.}
  \label{fig:afcs_model}
  \vspace{-.4cm}
\end{figure}

We apply the entire toolset on the Toyota Simulink model briefly discussed in Section~\ref{sec:eval} (part of the model is shown in Fig.~\ref{fig:afcs_model}).
First we run the translator:
\begin{lstlisting}[xleftmargin=.05\textwidth]
./simulink2isabelle.py afcs.slx -const -type real -iter -sim
\end{lstlisting}
Option \texttt{-iter} generates code that represents the behavior of the
system over time, and option \texttt{-sim} generates a Python simulation
script.

\begin{figure}
  \centering
  \includegraphics[scale=0.2]{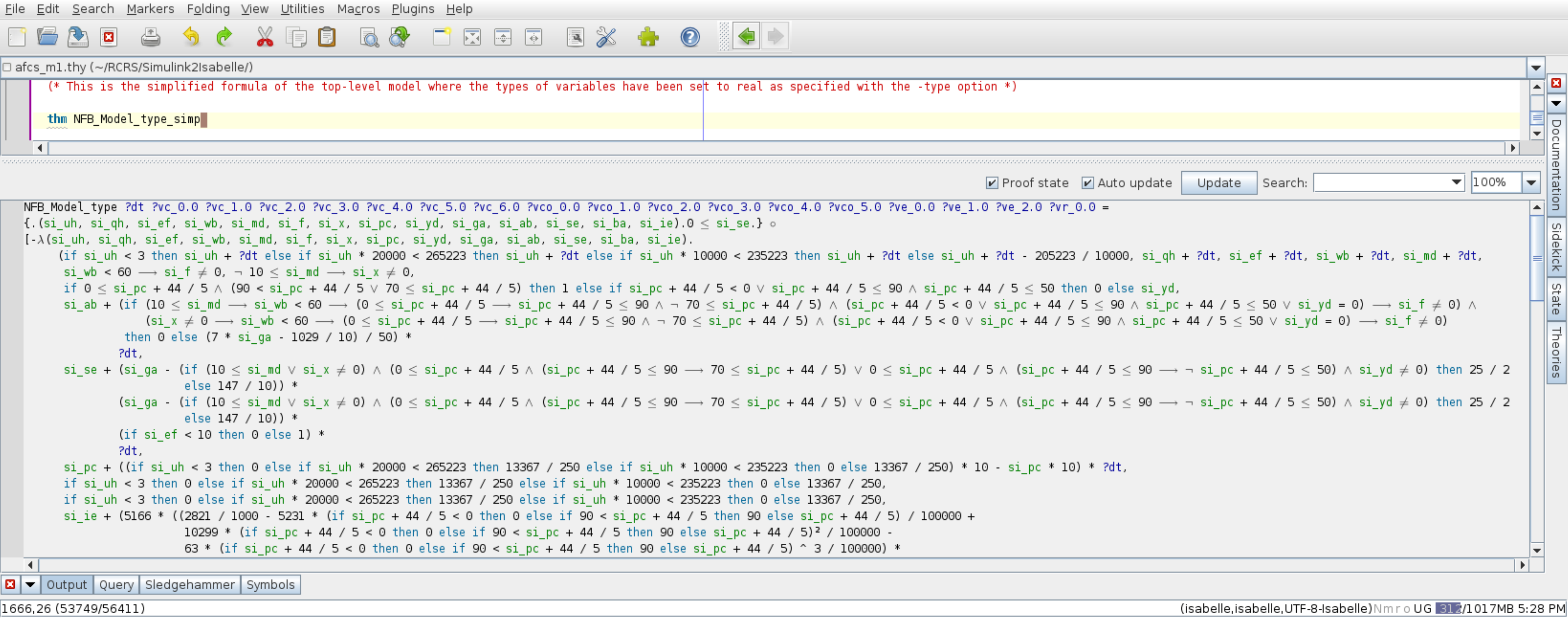}
  \caption{The simplified predicate transformer of the Fuel Control System.}
  \label{fig:afcs_contract}
  \vspace{-.2cm}
\end{figure}

The resulting file \texttt{afcs.thy} is quite long and its processing by Isabelle takes about 15 seconds. 
The Analyzer detects no incompatibilities and computes a simplified top-level atomic component whose description is 8338 characters long (an excerpt is shown in Fig.~\ref{fig:afcs_contract}). 
We explain this top-level component, which has a non-trivial automatically generated state condition. 
Then we run the generated Python simulation code, compare the simulation trajectories with those of Simulink, and find that they are essentially identical.


\begin{thebibliography}{10}

\bibitem{backwright:98}
R.-J. Back and J.~von Wright.
\newblock {\em {Refinement Calculus}}.
\newblock Springer, 1998.

\bibitem{Boulton:1992}
R.~J. Boulton, A.~Gordon, M.~J.~C. Gordon, J.~Harrison, J.~Herbert, and J.~V.
  Tassel.
\newblock Experience with embedding hardware description languages in {HOL}.
\newblock In {\em IFIP TC10/WG 10.2 Intl. Conf. on Theorem Provers in Circuit
  Design}, pages 129--156. North-Holland Publishing Co., 1992.

\bibitem{AlfaroHenzingerFSE01}
L.~de~Alfaro and T.~Henzinger.
\newblock Interface automata.
\newblock In {\em Foundations of Software Engineering (FSE)}. ACM Press, 2001.

\bibitem{dijkstra:75}
E.~Dijkstra.
\newblock {Guarded commands, nondeterminacy and formal derivation of programs}.
\newblock {\em Comm. ACM}, 18(8):453--457, 1975.

\bibitem{DragomirPT16}
I.~Dragomir, V.~Preoteasa, and S.~Tripakis.
\newblock {Compositional Semantics and Analysis of Hierarchical Block
  Diagrams}.
\newblock In {\em SPIN}, pages 38--56. Springer, 2016.

\bibitem{RCRS_Toolset_arxiv_2017}
I.~Dragomir, V.~Preoteasa, and S.~Tripakis.
\newblock {The Refinement Calculus of Reactive Systems Toolset}.
\newblock {\em CoRR}, abs/1710.08195:1--12, 2017.

\bibitem{RCRS_Toolset_figshare}
I.~Dragomir, V.~Preoteasa, and S.~Tripakis.
\newblock {The Refinement Calculus of Reactive Systems Toolset - Feb 2018}.
\newblock figshare. \url{https://doi.org/10.6084/m9.figshare.5900911}, Feb.
  2018.

\bibitem{JinDKUB14b}
X.~Jin, J.~V. Deshmukh, J.~Kapinski, K.~Ueda, and K.~Butts.
\newblock {Powertrain Control Verification Benchmark}.
\newblock In {\em Proceedings of the 17th International Conference on Hybrid
  Systems: Computation and Control}, HSCC'14, pages 253--262. ACM, 2014.

\bibitem{NipkowPW02}
T.~Nipkow, L.~C. Paulson, and M.~Wenzel.
\newblock {\em {Isabelle/HOL} --- A Proof Assistant for Higher-Order Logic}.
\newblock LNCS 2283. Springer, 2002.

\bibitem{PDT:2016}
V.~Preoteasa, I.~Dragomir, and S.~Tripakis.
\newblock A nondeterministic and abstract algorithm for translating
  hierarchical block diagrams.
\newblock {\em CoRR}, abs/1611.01337, 2016.

\bibitem{PDT17}
V.~Preoteasa, I.~Dragomir, and S.~Tripakis.
\newblock {The Refinement Calculus of Reactive Systems}.
\newblock {\em CoRR}, abs/1710.03979, 2017.

\bibitem{DragomirPreoteasaTripakisFORTE2017}
V.~Preoteasa, I.~Dragomir, and S.~Tripakis.
\newblock {Type Inference of Simulink Hierarchical Block Diagrams in Isabelle}.
\newblock In {\em FORTE}, 2017.

\bibitem{PreoteasaT14}
V.~Preoteasa and S.~Tripakis.
\newblock {Refinement calculus of reactive systems}.
\newblock In {\em Embedded Software (EMSOFT), 2014 International Conference
  on}, pages 1--10, Oct 2014.

\bibitem{TripakisLHL11}
S.~Tripakis, B.~Lickly, T.~A. Henzinger, and E.~A. Lee.
\newblock {A Theory of Synchronous Relational Interfaces}.
\newblock {\em ACM TOPLAS}, 33(4):14:1--14:41, July 2011.

\end{thebibliography}
\end{document}